# From Virtual Gains to Real Pains: Potential Harms of Immersive Exergames

Position Statement for the Workshop "Novel Approaches for Understanding and Mitigating Emerging New Harms in Immersive and Embodied Virtual Spaces"

SEBASTIAN CMENTOWSKI, HCI Games Group, Stratford School of Interaction Design and Business, University of Waterloo, Canada

SUKRAN KARAOSMANOGLU, Human-Computer Interaction, Universität Hamburg, Germany

FRANK STEINICKE, Human-Computer Interaction, Universität Hamburg, Germany

## POSITION STATEMENT

Digitalization and virtualization are parts of our everyday lives in almost all aspects ranging from work, education, and communication to entertainment. A novel step in this direction is the widespread interest in extended reality (XR) [2]. The newest consumer-ready head-mounted displays (HMD) such as Meta Quest 3 or Apple Vision Pro, have reached unprecedented levels of visual fidelity, interaction capabilities, and computational power. The built-in pass-through features of these headsets enable both virtual reality (VR) and augmented reality (AR) with the same devices. However, the immersive nature of these experiences is not the only groundbreaking difference from established forms of media.

XR technology offers unique interaction possibilities in virtual spaces. In contrast to other platforms, like PCs or tablets, XR builds heavily on spatial interactions with the virtual environment. Most current XR systems feature built-in six degrees of freedom (6-DOF) tracking of the HMD, controllers, or users' hands, allowing for natural and intuitive controls like grabbing or walking. This paradigm change towards physical movements not only boosts immersion and realism but also opens the possibility for entirely novel immersive experiences.

The spatial nature of XR has led to the use of these technologies for physical training and rehabilitation, often as so-called exergames [1, 2]. These games use the immersive power of XR technology, and feature physical exercises with engaging gameplay to improve exercise adherence and enjoyment. Today, exergames are among the best-selling XR applications, a trend that is highly positive, especially considering the World Health Organization (WHO) identified insufficient physical activity as an increasingly global problem [5]. As such, the captivating effect of immersive platforms offers an enormous potential to motivate people to exercise and thereby fight sedentary lifestyles. However, this also creates the risk of new (physical and social) harms that are not yet fully understood.

Like other forms of exercise, XR-based training also bears the risk of physical injury. For example, users could sustain injuries from over-exertion, wrong executions, or repetitive stress. However, we also see significant differences to traditional sports that may heighten these risks for XR exergames. Specifically, such applications are typically used in an unsupervised home setting. Also, the gamified nature of such experiences might further contribute, e.g., due to repetitive game mechanics or over-commitment due to rewards or social pressure. Similarly, due to the limited capacity of consumer-level headsets, current games do not track every part of the player's body and therefore may be offering exercise opportunities that are not entirely safe for players to perform. Lastly, since users often cannot see the real world while wearing XR headsets, this likely creates safety issues (e.g., collision).

Apart from the physical risk of injuries, immersive training experiences also bear the risk of only targeting the ideal, stereotypical user and, thereby, excluding other, potentially already marginalized, user groups. Without tailoring to individual persons, this approach could have serious concerns and consequences. For instance, on the one hand, it easily leads to people partaking in exercises that are not suited for their fitness level, causing either poor results (in case of under-exertion) or injuries (in case of over-exertion). On the other hand, users with limited abilities might not be able to consume these applications at all. With careful design, we could shape XR-based training into a new form of physical activity that is accessible and fun for everyone.

Given the technological developments, we believe targeting risk factors is of significant importance and relevance. We advocate for a broader discussion in the research community and call for more research to understand and resolve these risk factors for future movement-centered applications. Specifically, we are interested in the following topics:

- *Understanding Risks of XR Exergames*: While there has been individual research to create better, expert-driven exergames based on established training routines and to develop experiences for specific user groups with diverging needs, these are only individual cases. We are currently very interested in gaining a more holistic understanding of how users consume exergames, which injuries they may have sustained or barriers they have encountered, and if the gaming community has already established strategies to mitigate these risks and issues.
- *Creating Strategies and Recommendations to Improve XR-based Training*: Ultimately, we see the importance of articulating and formalizing strategies and recommendations for using XR applications for physical training. Formulating such guidelines would require a highly interdisciplinary team of experts and profound knowledge of the shortcomings of current games. However, creating those could be the first step in developing the next generation of safe and inclusive exergames that deliver sustainable and lasting health benefits for a growing audience.
- *Leveraging Artificial Intelligence to Create Personalized Experiences*: In the long term, we see the potential of using Artificial Intelligence (AI) to create better and more personalized training experiences. With more knowledge on training efficacy and risk factors, machine learning models could be key in realizing games that adapt dynamically to individual needs and progressions with longer use.

**Sebastian Cmentowski** is a postdoctoral fellow in the HCI Games Group at University of Waterloo, Canada. His PhD centered around improving the user experience in VR games by designing novel interaction concepts. Recently, he has been focusing on fighting sedentary lifestyles with immersive exergames. His work on designing VR exergames for jump training, co-authored with Sukran Karaosmanoglu, has received an honorable mention award at CHI 2023 [1].

**Sukran Karaosmanoglu** is a PhD student at the Human-Computer Interaction group, Universität Hamburg, Germany. For her PhD, Sukran works on creating accessible and health-focused XR experiences for various user groups. She has published multiple papers on immersive exergame design and game-based interventions for older adults [3, 4]. Another research focus is asymmetric games—games that bring together diverse user groups (e.g., young and old players). At CHI 2024, Sukran will present a comprehensive review of the current state of XR exergame research [2].

**Frank Steinicke** is a professor of Human-Computer Interaction at the Department of Informatics at the Universität Hamburg, Germany. His research interests are focused on understanding the human perceptual, cognitive, and motor abilities and limitations to improve interactions and experiences in computer-mediated realities. Frank Steinicke regularly serves as a panelist and speaker at major events in VR and human-computer interaction.